\documentstyle[aps,prd,eqsecnum,preprint]{revtex}
\begin{document}
\draft
\title{Search templates for gravitational waves from inspiraling binaries: \\
	Choice of template spacing}
\author{Benjamin J. Owen}
\address{ Theoretical Astrophysics, California Institute of Technology,
	Pasadena, California 91125 \\
	owen@tapir.caltech.edu}
\date{Submitted to Physical Review D: November 7, 1995}
\maketitle
\begin{abstract}

Gravitational waves from inspiraling, compact binaries will be searched for
in the output of the LIGO/VIRGO interferometric network by the method of
``matched filtering''---i.e., by correlating the noisy output of each
interferometer with a set of theoretical waveform templates.
These search templates will be a discrete subset of a continuous,
multiparameter family, each of which approximates a possible signal.
The search might be performed {\it hierarchically}, with a first pass
through the data using a low threshold and a coarsely-spaced, few-parameter
template set, followed by a second pass on threshold-exceeding data segments,
with a higher threshold and a more finely spaced template set that might have
a larger number of parameters.
Alternatively, the search might involve a single pass through the data using
the larger threshold and finer template set.
This paper extends and generalizes the Sathyaprakash-Dhurandhar (S-D)
formalism for choosing the discrete, finely-spaced template set used in the
final (or sole) pass through the data, based on the analysis of a single
interferometer.
The S-D formalism is rephrased in geometric language by introducing a metric
on the continuous template space from which the discrete template set
is drawn.
This template metric is used to compute the loss of signal-to-noise ratio
and reduction of event rate which result from the coarseness of the
template grid.
Correspondingly, the template spacing and total number ${\cal N}$ of templates
are expressed, via the metric, as functions of the reduction in event rate.
The theory is developed for a template family of arbitrary dimensionality
(whereas the original S-D formalism was restricted to a single nontrivial
dimension).
The theory is then applied to a simple post$^1$-Newtonian template family
with two nontrivial dimensions.
For this family, the number of templates ${\cal N}$ in the finely-spaced grid
is related to the spacing-induced fractional loss ${\cal L}$ of event rate
and to the minimum mass $M_{\min}$ of the least massive star in the binaries
for which one searches by
${\cal N} \sim 2\times 10^5(0.1/{\cal L})(0.2~M_\odot/M_{\min})^{2.7}$
for the first LIGO interferometers and by
${\cal N} \sim 8\times 10^6(0.1/{\cal L})(0.2~M_\odot/M_{\min})^{2.7}$
for advanced LIGO interferometers.
This is several orders of magnitude greater than one might have expected
based on Sathyaprakash's discovery of a near degeneracy in the parameter space,
the discrepancy being due to this paper's
lower choice of $M_{\min}$
and more stringent choice of ${\cal L}$.
The computational power ${\cal P}$
required to process the steady stream of incoming data
from a single interferometer through the closely-spaced set of templates
is given in floating-point operations per second by
${\cal P} \sim 2\times 10^{10}(0.1/{\cal L})(0.2~M_\odot/M_{\min})^{2.7}$
for the first LIGO interferometers and by
${\cal P} \sim 3\times 10^{11}(0.1/{\cal L})(0.2~M_\odot/M_{\min})^{2.7}$
for advanced LIGO interferometers.
This will be within the capabilities of LIGO-era computers,
but a hierarchical search may still be desirable
to reduce the required computing power.

\end{abstract}

\narrowtext

\section{Introduction}

Compact binary star systems are likely to be an important source
of gravitational waves for the broadband laser interferometric detectors
now under construction~\cite{Thorne 300yrs}, as they are the
best understood of the various types of postulated gravity wave sources
in the detectable frequency band and their waves should carry a large
amount of information.
Within our own galaxy, there are three known neutron star binaries whose
orbits will decay completely under the influence of gravitational radiation
reaction within less than one Hubble time, and it is almost certain that there
are many more as yet undiscovered.
Current estimates of the rate of neutron star/neutron star (NS/NS)
binary coalescences~\cite{Phinney,Narayan et al}
based on these (very few) known systems
project an event rate of three per year within a distance of roughly 200 Mpc;
and estimates based on the evolution of progenitor, main-sequence binaries
\cite{Tutukov&Yungelson}
suggest a distance of as small as roughly 70 Mpc for three events per year.
These distances correspond to a signal strength which is within the target
sensitivities of the LIGO and VIRGO interferometers~\cite{LIGO,VIRGO}.
However, to find the signals within the noisy LIGO/VIRGO data will require
a careful filtering of the interferometer outputs.
Because the predicted signal strengths
lie so close to the level of the noise, it will
be necessary to filter the interferometer data streams in order to detect
the inspiral events against the background of
spurious events generated by random noise.

The gravitational waveform generated by an inspiraling compact binary
has been calculated using a combination of post-Newtonian
and post-Minkowskian expansions~\cite{Blanchet&Damour 2PN,Will&Wiseman 2PN}
to post$^2$-Newtonian order by the consortium of Blanchet, Damour, Iyer,
Will, and Wiseman~\cite{2PN consortium}, and will be calculated to
post$^3$-Newtonian order long before the LIGO and VIRGO interferometers
come on-line (c.~2000).
Because the functional form of the expected signal is so well-known, it is
an ideal candidate for matched filtering, a venerable and widely known
technique which is laid out in detail elsewhere~\cite{matched filtering}
and briefly summarized here:

The matched filtering strategy is to compute a
cross-correlation between the interferometer output and a template waveform,
weighted inversely by the noise spectrum of the detector.
The signal-to-noise ratio is defined as the value of the cross-correlation
of the template with a particular stretch of data
divided by the rms value of the cross-correlation of the template with pure
detector noise.
If the signal-to-noise ratio exceeds a certain threshold,
which is set primarily to control the
rate of false alarms due to fluctuations of the noise,
a detection is registered.
If the functional form of the template is identical to that of the signal,
the mean signal-to-noise ratio in the presence of a signal
is the highest possible for any linear data processing technique,
which is why matched filtering is also known as optimal filtering.

In practice, however, the template waveforms will differ somewhat from
the signals.
True gravitational-wave signals from inspiraling binaries
will be exact solutions to the Einstein
equations for two bodies of non-negligible mass,
while the templates used to search for these signals will be, at best,
finite-order approximations to the exact solutions.
Also, true signals will be characterized by many parameters
(e.g. the masses of the two objects, their spins, the eccentricity and
orientation of the orbit...),
some of which might be neglected in construction of the search templates.
Thus, the true signals will lie somewhat outside
the submanifold formed by the search templates in the full manifold
of all possible detector outputs (see Fig.~\ref{fig:manifold}).

Apostolatos~\cite{Apostolatos} has defined the ``fitting factor'' $FF$
to quantitatively describe the closeness of the true signals to the
template manifold in terms of the reduction of the
signal-to-noise ratio
due to cross-correlating a signal lying outside the manifold
with all the templates lying inside the manifold.
If the fitting factor of a template family is unity,
the signal lies in the template manifold.
If the fitting factor is less than unity,
the signal lies outside the manifold,
and the fitting factor represents the cross-correlation
between the signal and the template nearest it in the template manifold.

Even if the signal were to lie within the template manifold,
it would not in general correspond to any of the actual templates used to
search the data.
The parameters describing the search templates (masses, spins, etc.)
can vary continuously throughout a finite range of values.
The set of templates characterized by the continuously varying
parameters is of course infinite, so
the interferometer output must be cross-correlated with a finite subset of
the templates whose parameter values vary in discrete steps from one
template to the next.
This subset (the ``discrete template family'') has measure zero
on the manifold of the full set of possible templates
(the ``continuous template family''), so the template which
most closely matches a signal will generally
lie in between members of the discrete template family
(again, see Fig.~\ref{fig:manifold}).
The mismatch between the signal and the nearest of the discrete templates
will cause some reduction in the signal-to-noise ratio and therefore in the
observed event rate,
as some signals which would lie above the threshold if cross-correlated with
a perfectly matched filter are driven below the threshold by the mismatch.
Thus the spacing between members of the discrete template family must be chosen
so as to render acceptable the loss in event rate,
without requiring a prohibitive amount of computing power
to numerically perform
the cross-correlations of the data stream with all of the discrete templates.

The high computational demands
of a laser interferometric detector may in fact
make it desirable to perform a {\it hierarchical search}.
In a hierarchical search, each stretch of data is first
filtered by a set of templates which rather sparsely populates the
manifold,
and stretches which fail to exceed a relatively low signal-to-noise threshold
are discarded.
The surviving stretches of data are filtered by a larger set of templates
which more densely populates the manifold,
and are subjected to a higher threshold.
The spareseness of the first-pass template set
insures that most of the data need only be filtered by a small number of
templates,
while the high threshold of the final pass
reduces the false alarm rate to an acceptable level.

Theoretical foundations for choosing the discrete set of templates
from the continuous family were laid by Sathyaprakash and Dhurandhar
for the case of white noise in Ref.~\cite{S&D 1}, and
for (colored) power-recycling interferometer
noise in Ref.~\cite{S&D 2}.
Both papers used a simplified (so-called ``Newtonian'')
version of the waveform which can be
characterized by a single parameter, the binary's ``chirp mass'' ${\cal M}$.
Recently, Sathyaprakash~\cite{Sathya note} began consideration of an improved,
``post-Newtonian'' set of templates characterized by two mass parameters.
He found that, by a judicious choice of the two parameters,
the spacing between templates can be made constant
in both dimensions of the intrinsic parameter space.
Sathyaprakash's parameters also make it obvious
(by producing a very large spacing in one of the dimensions)
that a two-parameter set of templates can be constructed
which, if it does not populate the manifold too densely,
need not be much more numerous than the one-parameter set of templates
used in Refs.~\cite{S&D 1,S&D 2}.

In this paper I shall recast the S-D formalism in geometric language which,
I believe, simplifies and clarifies the key ideas.
I shall also generalize the S-D formalism to an arbitrary spectrum
of detector noise and to a set of template shapes characterized by
more than one parameter.
This is necessary because, as Apostolatos~\cite{Apostolatos} has shown,
no one-parameter set of templates can be used to filter a
post-Newtonian signal without causing an unacceptably large loss of
signal-to-noise ratio.

In one respect, my analysis will be more specialized than that of
the S-D formalism.
My geometric analysis requires that the templates of the discrete set be
spaced very finely in order that certain analytical approximations may
be made, while the numerical methods of Sathyaprakash and Dhurandhar are
valid even for a large spacing between templates (as would be the case in
the early stages of a hierarchical search).
The small spacing approximation is justified on the grounds that at some
point, even in a hierarchical search, the data must be filtered by many
closely spaced templates in order to detect a reasonable fraction
(of order unity) of the binary inspirals occurring in the universe
within range of the LIGO/VIRGO network.

The rest of this paper is organized as follows:
In Sec.~\ref{sec:formalism},
I develop my generalized, geometric variant of the S-D formalism.
I then apply this formalism
to the general problem of detection of gravitational waves
from inspiraling binaries, and develop general formulas for choosing
a discrete template family from a given continuous template family.
In Sec.~\ref{sec:1PN}, I detail an example of the use of my formalism,
choosing discrete templates from a continuous template family
which describes nonspinning, circularized binaries
to post$^1$-Newtonian order in the evolution of the waveform's phase.
I also estimate the computing power required for a single-pass
(non-hierarchical) search using this discrete template family,
and compare to the previous work of Sathyaprakash~\cite{Sathya note}.
Finally, in Sec.~\ref{sec:conclusion}, I summarize my results and
suggest future directions for research on the
choice of discrete search templates.

\section{Theory of Mismatched Filtering}
\label{sec:formalism}

In this section, a geometric, multiparameter
variant of the S-D formalism is developed.
Unless otherwise stated, the following conventions and definitions are assumed:

Following Cutler and Flanagan~\cite {Cutler&Flanagan},
we define the inner product between two functions of time
$a(t)$ and $b(t)$
(which may be templates or interferometer output) as
\begin{eqnarray}
\label{inner product definition}
\langle a | b \rangle &\equiv&
2 \int_{0}^\infty df
\frac{\tilde a^*(f)\tilde b(f) + \tilde a(f)\tilde b^*(f)}{S_h(f)} \nonumber \\
&=& 4\Re\left[
\int_0^\infty df \frac{\tilde a^*(f)\tilde b(f)}{S_h(f)}\right] .
\end{eqnarray}
Here $\tilde a(f)$ is the Fourier transform of $a(t)$,
\begin{equation}
\tilde a(f)\equiv\int_{-\infty}^{\infty}dt~e^{i2\pi ft}a(t),
\end{equation}
and $S_h(f)$ is the detector's noise spectrum, defined below.

The interferometer output $o(t)$ consists of noise $n(t)$ plus a signal
${\cal A}s(t)$, where
${\cal A}$ is a dimensionless, time-independent amplitude and
$s(t)$ is normalized such that
$\langle s | s \rangle = 1$.
Thus, ${\cal A}$ describes the strength of a signal
and $s(t)$ describes its shape.

Waveform templates are denoted by $u(t;\bbox{\mu},\bbox{\lambda})$,
where $\bbox{\lambda}$ is the vector of ``intrinsic'' or ``dynamical''
parameters characterizing the template shape and
$\bbox{\mu}$ is the vector of ``extrinsic'' or ``kinematical'' parameters
describing the offsets of the waveform's endpoint.
Examples of intrinsic parameters $\lambda^i$ are the masses and spins
of the two objects in a compact binary;
examples of extrinsic parameters $\mu^i$ are the time of a compact binary's
final coalescence $t_0$ and the phase of the waveform at coalescence $\Phi_0$.

Templates are assumed to be normalized such that
$\langle u(\bbox{\mu},\bbox{\lambda})|u(\bbox{\mu},\bbox{\lambda})\rangle =1$
for all $\bbox{\mu}$ and $\bbox{\lambda}$.

Expectation values of various quantities over an infinite ensemble of
realizations of the noise are denoted by ${\rm E}[~]$.

The interferometer's strain spectral noise density
$S_h(f)$ is the one-sided spectral density, defined by
\begin{equation}
{\rm E}[\tilde n(f_1)\tilde n^*(f_2)] =
\frac{1}{2}\delta (f_1 - f_2) S_h(f_1)
\end{equation}
for positive frequencies and undefined for negative frequencies.
The noise is assumed to have a Gaussian probability distribution.

Newton's gravitational constant $G$ and the speed of light $c$ are set
equal to one.

\subsection{Formalism}

In developing our formalism, we begin by defining the signal-to-noise ratio.
For any single template $u(t)$ of unit norm,
the cross-correlation with pure noise $\langle n | u \rangle$ is a
random variable with mean zero and variance unity (cf.
Sec.~II.B. of Ref.~\cite{Cutler&Flanagan}, wherein it is shown that
${\rm E}[\langle n | a\rangle\langle n | b\rangle] = \langle a | b\rangle$).
The signal-to-noise ratio of a given stretch of data $o(t)$,
after filtering by $u(t)$, is defined to be
\begin{equation}
\label{S/N definition}
\rho \equiv \frac{\langle o | u\rangle}{{\rm rms}~\langle n | u\rangle} =
\langle o | u\rangle .
\end{equation}
This ratio
is the statistic which is compared to a predetermined threshold
to decide if a signal is present.

If the template $u$ is the same as the signal $s$,
it optimally filters the signal,
and the corresponding (mean) optimal signal-to-noise ratio is
\begin{eqnarray}
\label{optimal S/N}
{\rm E}[\rho] &=& {\rm E}[\langle n+{\cal A}u | u\rangle] \nonumber \\
&=& {\cal A}.
\end{eqnarray}
If the template $u$ used to filter the data
is not exactly the same as the signal $s$,
the mean signal-to-noise ratio is decreased somewhat
from its optimal value:
\begin{eqnarray}
\label{non-optimal S/N}
{\rm E}[\rho] &=& {\rm E}[\langle n+{\cal A}s | u\rangle] \nonumber \\
&=& {\cal A}\langle s | u\rangle.
\end{eqnarray}
The inner product $\langle s | u \rangle$,
which is bounded between zero and one, is
the fraction of the optimal ${\rm E}[\rho]$ retained
in the mismatched filtering case, and as such is
a logical measure of the effectiveness
of the template $u$ in searching for the signal shape $s$.

Now suppose that we search for the signal with a family of templates
specified by an extrinsic parameter vector $\bbox{\mu}$
and an intrinsic parameter vector $\bbox{\lambda}$.
Let us denote the values of the parameters of the actual templates
by $(\bbox{\mu}_{(k)},\bbox{\lambda}_{(k)})$.
For example, $\mu^1_{(k)}$ might be the value of the time
$t_0$ of coalescence for the $k$th template in the family,
and $\mu^2_{(k)}$ might be the phase of the $k$th template waveform
at coalescence.

The search entails computing, via fast Fourier transforms (FFT's),
all the inner products
$\langle o | u(\bbox{\mu}_{(k)},\bbox{\lambda}_{(k)}) \rangle$
for $k = 1, 2, \ldots\;$
In these numerical computations, the key distinction between
the extrinsic parameters $\bbox{\mu}$
and the intrinsic parameters $\bbox{\lambda}$ is this:
One explores the whole range of values of $\bbox{\mu}$
very quickly, automatically, and efficiently for a fixed value of
$\bbox{\lambda}$;
but one must do these explorations separately for each of the
$\bbox{\lambda}_{(k)}$.
In this sense, dealing with the extrinsic parameters is far easier
and more automatic than dealing with the intrinsic ones.

As an example (for further detail see Sec.\ 16.2.2 of Schutz~\cite{Schutz}),
for a given stretch of data one explores {\it all} values of
the time of coalescence $(t_0\equiv\mu^1)$ of a compact binary
simultaneously (for fixed values of the other template parameters)
via a single FFT.
If we write the Fourier transform (for notational simplicity) as
a continuous integral rather than a discrete sum, we get
\begin{equation}
\renewcommand{\arraystretch}{1.5}
\begin{array}{l}
\langle o | u(\bbox{\mu}_{(k)},\bbox{\lambda}_{(k)}) \rangle \\
\quad=\int_0^\infty df~e^{i2\pi ft_0}\tilde{o}^*(f)
\tilde{u}(f;\mbox{other }\bbox{\mu}_{(k)},\bbox{\lambda}_{(k)}) .
\end{array}
\end{equation}
The discrete FFT yields the discrete analog
of the function of $t_0$ as shown above,
an array of numbers containing the values of the Fourier transform
for all values of $t_0$.

Because, for fixed $\bbox{\lambda}_{(k)}$,
the extrinsic parameters $\bbox{\mu}$ are dealt with
so simply and quickly in the search,
throughout this paper we shall focus primarily on a template family's
intrinsic parameters $\bbox{\lambda}$,
which govern the shape of the template.
Correspondingly, we shall adopt the following quantity
as our measure of the effectiveness with which a particular
template shape---i.e.\ a particular vector $\bbox{\lambda}_{(k)}$
of the intrinsic parameters---matches the incoming signal:
\begin{equation}
\label{sort of match}
\renewcommand{\arraystretch}{.6}
\begin{array}[t]{c}{\textstyle\max}\\{\scriptstyle\mu}\end{array}
\langle s | u(\bbox{\mu},\bbox{\lambda}_{(k)}) \rangle .
\end{equation}
Here the maximization is over all continuously varying
values of the extrinsic parameters.
Then the logical measure of the effectiveness
of the entire discrete family of templates
in searching for the signal shape is
\begin{equation}
\renewcommand{\arraystretch}{.6}
\begin{array}[t]{c}{\textstyle\max}\\{\scriptstyle k}\end{array}
\bigl[
\begin{array}[t]{c}{\textstyle\max}\\{\scriptstyle\mu}\end{array}
\langle s | u(\bbox{\mu},\bbox{\lambda}_{(k)}) \rangle
\bigr] ,
\end{equation}
which is simply~(\ref{sort of match})
maximized over all the discrete template shapes.

In order to focus on the issue of discretization
of the template parameters rather than
on the inadequacy of the continuous template family,
let us assume that the signal shape $s$ is identical to some template.
The discussion of the preceding paragraphs
suggests that in discussing the discretization of the template parameters
we will want to make use of the {\it match} between
two templates $\tilde u(f;\bbox{\mu},\bbox{\lambda})$
and $\tilde u(f;\bbox{\mu}+\Delta\bbox{\mu},
\bbox{\lambda}+\Delta\bbox{\lambda})$
which we will define as
\begin{equation}
\label{match definition}
M(\bbox{\lambda},\Delta\bbox{\lambda})\equiv
\renewcommand{\arraystretch}{.6}
\begin{array}[t]{c}{\textstyle\max}\\{\scriptstyle\mu,\Delta\mu}\end{array}
\langle u(\bbox{\mu},\bbox{\lambda}) |
u(\bbox{\mu}+\Delta\bbox{\mu},\bbox{\lambda}+\Delta\bbox{\lambda}) \rangle .
\end{equation}
This quantity, which is known in the theory of hypothesis testing
as the {\it ambiguity function,} is the
fraction of the optimal signal-to-noise ratio obtained by
using a template with intrinsic parameters $\bbox{\lambda}$
to filter a signal identical in shape to a template with
intrinsic parameters $\bbox{\lambda}+\Delta\bbox{\lambda}$.

Using the match~(\ref{match definition})
it is possible to quantify our intuitive notion of how
``close'' two template shapes are to each other.
Since the match clearly has a maximum value
of unity at $\Delta\bbox{\lambda}=0$,
we can expand in a power series about $\Delta\bbox{\lambda}=0$ to obtain
\begin{equation}
\label{match expansion}
M(\bbox{\lambda},\Delta\bbox{\lambda})\approx 1 +
\frac{1}{2}
\left(
\frac{\partial^2 M}{\partial\Delta\lambda^i \partial\Delta\lambda^j}
\right)
{\atop\scriptstyle\Delta\lambda^k = 0}
\Delta\lambda^i \Delta\lambda^j .
\end{equation}
This suggests the definition of a metric
\begin{equation}
\label{metric definition}
g_{ij}(\bbox{\lambda}) = -\frac{1}{2}\left(
\frac{\partial^2 M}{\partial\Delta\lambda^i \partial\Delta\lambda^j}\right)
{\atop\scriptstyle\Delta\lambda^k = 0}
\end{equation}
so that the {\it mismatch} $1-M$ between two nearby templates is equal to
the square of the proper distance between them:
\begin{equation}
\label{mismatch}
1-M=g_{ij}\Delta\lambda^i\Delta\lambda^j .
\end{equation}

Having defined a metric on the intrinsic parameter space,
we can now use it to calculate the spacing of the discrete template family
required to retain a given fraction of the ideal event rate.
Schematically, we can think of the templates as forming a lattice in the
$N$-dimensional intrinsic parameter space whose unit cell is an
$N$-dimensional hypercube with sides of proper length $dl$.
The worst possible case (lowest ${\rm E}[\rho ]$)
occurs if the point $\tilde{\bbox{\lambda}}$
describing the signal is exactly in the middle of one of the hypercubes.
If the templates are closely spaced, i.e.\ $dl\ll 1$,
such a signal has a squared proper distance
\begin{equation}
g_{ij}\Delta\lambda^i\Delta\lambda^j = N (dl/2)^2
\end{equation}
from the templates at the corners of the hypercube.

We define the {\it minimal match} $MM$ to be the match between
the signal and the nearest templates in this worst possible case,
i.e.\ the fraction of the optimal signal-to-noise ratio retained by
a discrete template family when the signal falls exactly
``in between'' the nearest templates.
This minimal match
is the same quantity that Dhurandhar and Sathyaprakash in Ref.~\cite{S&D 2}
denote as $\kappa^{-1}$;
but since it is the central quantity governing template spacing
it deserves some recognition in the form of its own name.
Our choice of name closely parallels the term ``fitting factor'' $FF$,
which Apostolatos introduced in Ref.~\cite{Apostolatos}
to measure the similarity between actual signals
and a continuous template family.

The minimal match, which is chosen by the experimenter
based upon what he or she considers
to be an acceptable loss of ideal event rate,
will determine our choice of spacing of the discrete template parameters
and therefore the number of discrete templates in the family.
More specifically, the experimenter will choose some desired value
of the minimal match $MM$;
and then will achieve this $MM$ by selecting the templates to reside
at the corners of hypercubes with edge $dl$ given by
\begin{equation}
\label{MM vs dl}
MM = 1 - N (dl/2)^2 .
\end{equation}
The number of templates in the resulting discrete template family will be
the proper volume of parameter space
divided by the proper volume per template $dl^N$, i.e.
\begin{equation}
\label{cal N definition}
{\cal N} = \frac{\int d^N\bbox{\lambda} \sqrt{{\rm det}~\|g_{ij}\|} }
{\left(2\sqrt{\left(1 - MM\right)/N}\right)^N}.
\end{equation}

\subsection{Inspiraling Binaries Detected by LIGO}

The formalism above applies to the detection of any set of signals
which have a functional form that depends on
a set of parameters which varies continuously over some range.
We now develop a more explicit formula for the metric, given an
analytical approximation to the LIGO noise curve and a particular
class of inspiraling binary signals.

We approximate the ``initial'' and ``advanced'' benchmark
LIGO noise curves by the following
analytical fit to Fig.~7 of Ref.~\cite{LIGO}:
\begin{equation}
\label{LIGO noise curve}
S_h(f) = \left\{
\begin{array}{ll}
\frac{1}{5}S_0\{(f/f_0)^{-4}+2[1+(f/f_0)^2]\},&f>f_s\\
\infty,&f<f_s,
\end{array}
\right.
\end{equation}
where $f_0$ is the ``knee frequency'' or frequency at which the
interferometer is most sensitive
(which is determined by the reflectivities of the mirrors and
is set by the experimenters to the frequency where
photon shot noise begins to dominate the spectrum)
and $S_0$ is a constant whose value is not important for our purposes.
This spectrum describes photon shot noise in the ``standard recycling''
configuration of the interferometer (second term)
superposed on thermal noise
in the suspension of the test masses (first term),
and it approximates seismic noise by setting $S_h$ infinite at frequencies
below the ``seismic-cutoff frequency'' $f_s$.

Throughout the rest of this paper,
the ``first LIGO noise curve'' will refer to (\ref{LIGO noise curve})
with $f_s$ = 40 Hz and $f_0$ = 200 Hz,
and the ``advanced LIGO noise curve'' will refer to (\ref{LIGO noise curve})
with $f_s$ = 10 Hz and $f_0$ = 70 Hz.
These numbers are chosen to closely fit Fig.~7 of Ref.~\cite{LIGO}
for the first LIGO interferometers and for the advanced LIGO benchmark.
In this paper, when various quantities
(such as the number of discrete templates)
are given including a scaling with $f_0$,
this indicates how the quantity changes
while $f_0$ is varied but $f_s/f_0$ is held fixed.

At this point it is useful to define
the moments of the noise curve~(\ref{LIGO noise curve}), following
Poisson and Will~\cite{Poisson&Will}, as
\begin{mathletters}
\label{moment definition}
\begin{eqnarray}
I(q) &\equiv&
S_h(f_0)\int_{f_s/f_0}^{f_c/f_0}dx\frac{x^{-q/3}}{S_h(x~f_0)} \nonumber \\
&=&\int_{f_s/f_0}^{f_c/f_0}dx\frac{5x^{-q/3}}{x^{-4}+2(1+x^2)}, \\
J(q) &\equiv& I(q)/I(7) .
\end{eqnarray}
The upper limit of integration $f_c$ denotes the coalescence frequency
or high-frequency cutoff of whatever template we are dealing with,
which very roughly corresponds to the last stable circular orbit of a test
particle in a non-spinning black hole's Schwarzschild geometry.
\end{mathletters}

For both first and advanced LIGO noise curves,
the majority of inspiraling binary search templates will occupy
regions of parameter space for which $f_c$ is many times $f_0$.
Because we will always be dealing with $I(q)$ for $q > 0$, and because
the noise term in the denominator of the integrand in
Eq.~(\ref{moment definition}) rises as
$f^2$ for $f \gg f_0$, we can simplify later calculations by approximating
$f_c = \infty$ in the definition of the moments.

To illustrate the metric formalism,
we shall use templates based on
a somewhat simplified version of the post-Newtonian expansion.
Since the inner product (\ref{inner product definition})
has negligible contributions from frequencies at which
the integrand oscillates rapidly, it is far more important to get the phase
of $\tilde u(f)$ right than it is to get the amplitude dependence.
Therefore,
we adopt templates based on the ``restricted'' post-Newtonian
approximation in which one discards all multipolar components
except the quadrupole,
but keeps fairly accurate track of the quadrupole component's phase
(for more details see
Secs.~II.C. and III.A. of Ref.~\cite{Cutler&Flanagan}).
Applying the stationary phase approximation to that quadrupolar waveform,
we obtain
\begin{equation}
\label{template definition}
\tilde u(f;\bbox{\lambda},\bbox{\mu}) =
f^{-7/6}\exp~i\left[-\frac{\pi}{4} - \Phi_0 + 2\pi f t_0 +
\Psi(f;\bbox{\lambda})\right],
\end{equation}
up to a multiplicative constant
which is set by the condition $\langle u | u \rangle = 1$
\cite{phase footnote}.

The function $\Psi$, describing the phase evolution in
(\ref{template definition}),
is currently known to post$^2$-Newtonian order
for the case of two nonspinning point masses in a circular orbit
about each other as
\begin{equation}
\label{2PN Psi}
\renewcommand{\arraystretch}{1.5}
\begin{array}{c}
\Psi(f;M,\eta)=\frac{3}{128}\eta^{-1}(\pi Mf)^{-5/3}\nonumber\\
\times\biggl[1+\frac{20}{9}\bigl(\frac{743}{336}+\frac{11}{4}\eta\bigr)
(\pi Mf)^{2/3}-16\pi(\pi Mf)\biggr.\nonumber\\
\biggl.+10\bigl(\frac{3058673}{1016064}+\frac{5429}{1008}\eta
+\frac{617}{144}\eta^2\bigr)(\pi Mf)^{4/3}\biggr]
\end{array}
\end{equation}
(cf. Eq.~(3.6) of Ref.~\cite{Poisson&Will}).
Here the mass parameters have
been chosen to be $M$, the total mass of the system, and $\eta$, the ratio
of the reduced mass to the total mass.

The actual amplitude ${\cal A}$ of a waveform is proportional to $1/R$,
where $R$ is the distance to the source;
and this lets us find the relation between minimal match and event rate
which we will need in order to wisely choose the minimal match.
Assuming that compact binaries are uniformly distributed throughout space
on large distance scales, this means that the rate of events
with a given set of intrinsic parameters and with an
amplitude greater than ${\cal A}$ is proportional to $1/{\cal A}^3$.
Setting a signal-to-noise threshold $\rho_0$ is equivalent to setting
a maximum distance $R_0 \propto 1/{\cal A}$
to which sources with a given set of intrinsic
parameters can be detected.
Thus, if we could search for signals with the entire continuous template
family, we would expect the observed event rate to scale as $1/\rho_0^3$.
This ideal event rate is an upper limit on what we can expect with a real,
discrete template family.

We can obtain a lower bound on the observed event rate by considering what
happens if all signals conspire to have parameters lying exactly in
between the nearest search templates.
In this case, all events will have reduced signal-to-noise ratios of
$MM$ times the optimal signal-to-noise ratio ${\cal A}$.
This is naively equivalent to optimally filtering with a threshold
of $\rho_0/MM$, so a pessimistic guess is
\begin{equation}
\label{ER vs MM}
\mbox{event rate }\propto\left(\frac{MM}{\rho_0}\right)^3
\end{equation}
for a fixed rate of false alarms.

In real life, $\rho_0$ is affected by the total number of discrete templates
and by the minimal match of the discrete template family.
This can be seen by the fact that the signal-to-noise ratio,
\begin{equation}
\rho\equiv
\renewcommand{\arraystretch}{.6}
\begin{array}[t]{c}{\textstyle\max}\\{\scriptstyle k}\end{array}
\bigl[
\begin{array}[t]{c}{\textstyle\max}\\{\scriptstyle\mu}\end{array}
\langle o | u(\bbox{\mu},\bbox{\lambda}_{(k)}) \rangle
\bigr] ,
\end{equation}
is the maximum of a number of random variables.
The covariance matrix of these variables will be determined by
the minimal match,
and will itself determine the probability distribution of $\rho$
(in the absence of a signal)
which is used to set the threshold $\rho_0$ in order to
keep the false alarm rate below a certain level.
However, since these effects are fairly small at high signal-to-noise ratios
(such as those to be used by LIGO) and the issue of choosing thresholds
is a problem worthy of its own paper \cite{thresholds},
we will use (\ref{ER vs MM}) for the rest of this paper.

For this two-parameter template family,
the formula for the match (\ref{match definition}) can be simplified somewhat
by explicitly performing the maximization over the extrinsic parameters
$\bbox{\mu}$ and $\Delta\bbox{\mu}$.
Since the integrand in the inner product~(\ref{inner product definition})
depends on $\bbox{\mu}$ and $\Delta\bbox{\mu}$
as $\exp~i[2\pi f\Delta t_0 - \Delta\Phi_0]$,
there is no dependence on $\bbox{\mu}$ but only on $\Delta\bbox{\mu}$.
Maximizing over $\Delta\Phi_0$ is easy: instead of taking the real
part of the integral in the inner product (\ref{inner product definition}),
we take the absolute value.

To maximize over $\Delta t_0$ we go back a step.
Let us define $\lambda^0 \equiv t_0$, and consider the $(N+1)$-dimensional
space formed by $\lambda^0$ and $\lambda^j$.
We expand the inner product between adjacent templates
to quadratic order in $\Delta\lambda^\alpha$
to get a preliminary metric $\gamma_{\alpha\beta}$,
where the Greek indices range from $0$ to $N$ (Latin indices range from
$1$ to $N$):
\widetext
\begin{equation}
\label{premetric-def}
\gamma_{\alpha\beta}(\bbox{\lambda}) = -\frac{1}{2}
\left[
\frac{\partial^2}{\partial\Delta\lambda^\alpha\partial\Delta\lambda^\beta}
\left\{\frac{\left|\int_0^{\infty}df
\frac{\displaystyle f^{-7/3}}{\displaystyle S_h(f)}
\exp~i[2\pi f\Delta t_0+
\Delta\Psi(f;\lambda^j,\Delta\lambda^j)]\right| }
{\int_0^{\infty}df\frac{\displaystyle f^{-7/3}}{\displaystyle S_h(f)}}\right\}
\right]
{\atop\Delta\lambda^\alpha = 0}.
\end{equation}
Here $\Delta\Psi \equiv \Psi(f;\lambda^j+\Delta\lambda^j)- \Psi(f;\lambda^j)$
\cite{gamma footnote}.
\narrowtext

We define the moment functional ${\cal J}$ such that, for a function $a$,
\begin{equation}
\label{functional definition}
{\cal J}[a] =
\frac{1}{I(7)}\int_{f_s/f_0}^{f_c/f_0} dx\frac{x^{-7/3}}{S_h(x~f_0)}a(x),
\end{equation}
and thus
\begin{equation}
\label{series functional}
{\cal J}\left[\sum_n a_n x^n\right] = \sum_n a_n J(7-3n).
\end{equation}
We also define the quantities $\psi_\alpha$ such that
\begin{equation}
\label{psi-def}
\psi_0 \equiv 2\pi f,\quad
\psi_j \equiv \frac{\partial\Delta\hat{\Psi}}{\partial\Delta\lambda^j},
\end{equation}
where the derivative is evaluated at $\Delta\lambda^j = 0$ and
$\hat{\Psi}$ is the part of $\Psi$ in Eq.~(\ref{template definition})
that is frequency dependent (any non-frequency-dependent,
additive parts of
$\Psi$ are removed when we take the absolute value in the maximized
inner product).
Evaluation of the derivative in Eq.~(\ref{premetric-def}) then shows that,
in the limit $f_c/f_0\to \infty$,
\begin{equation}
\label{premetric}
\gamma_{\alpha\beta} = \frac{1}{2}
\left({\cal J}[\psi_\alpha\psi_\beta] -
{\cal J}[\psi_\alpha]{\cal J}[\psi_\beta]\right).
\end{equation}

Finally, we minimize
$\gamma_{\alpha\beta}\Delta\lambda^\alpha\Delta\lambda^\beta$
with respect to $\Delta t_0$
(i.e., we project $\gamma_{\alpha\beta}$
onto the subspace orthogonal to $t_0$)
and thereby obtain the following expression
for the metric of our continuous template family:
\begin{equation}
\label{metric conversion}
g_{ij} = \gamma_{ij} - \frac{\gamma_{0i}\gamma_{0j}}{\gamma_{00}}.
\end{equation}
By taking the square root of the determinant of this metric
and plugging it into
Eq.~(\ref{cal N definition}),
we can compute the number of templates ${\cal N}$ that we need in our
discrete family as a function of our desired minimal match $MM$,
or equivalently of the loss of ideal event rate.

\section{Example: Circularized Nonspinning Binaries to
Post$^1$-Newtonian Order}
\label{sec:1PN}

Although the phase of the inspiraling binary signal has recently been
calculated to post$^2$-Newtonian order~\cite{2PN consortium},
it is useful to calculate the number of templates that would be
required in a universe where the waveforms evolve only to
post$^1$-Newtonian order
and all binaries are composed of nonspinning objects in circular orbits.
There are several reasons for this exercise.
\begin{enumerate}
\item
Apostolatos~\cite{Apostolatos} has shown that amplitude modulation of the
waveform due to spin effects is important in an inspiraling binary search
only for a few extremal combinations of parameters,
and also that (at higher post-Newtonian order)
templates without spin-related phase modulation can
match phase modulated signals almost as well
as can templates that include spin parameters.
Therefore the bulk of the final set of templates actually used when the
detectors come on-line will not need to include the extra spin
parameters, and we may ignore them in this preliminary work.
\item
We assume circular orbits because
gravitational radiation reaction circularizes most eccentric orbits
on a timescale much less than the lifetime
of a compact binary~\cite{circular orbits}.
\item
The phase of the templates is truncated at post$^1$-Newtonian order
for simplicity.
Although Apostolatos has demonstrated in Ref.~\cite{Apostolatos} that
post$^1$-Newtonian templates will not have a large enough fitting factor
to be useful, consideration of such a set is a first step toward
obtaining an adequate set of templates---and
it is a particularly important step since the metric coefficients will
turn out to be constant over the template manifold.
\end{enumerate}

\subsection{Calculation of the 2-Dimensional Metric}

Having chosen as the continuous template family
the set of post$^1$-Newtonian, circular, spinless binary waveforms,
we must now choose the discrete templates
from within this continuous family.
The first step is to calculate the coefficients of the metric
on the two-dimensional dynamical parameter space.

It is convenient to
change the mass parameterization from the variables $(M,\eta)$
to the Sathyaprakash variables~\cite{Sathya note}
\begin{eqnarray}
\tau_1 &=& \frac{5}{256}\eta^{-1}M^{-5/3}(\pi f_0)^{-8/3},\label{def:tau1} \\
\tau_2 &=& \frac{5}{192}(\eta M)^{-1}
\left(\frac{743}{336}+\frac{11}{4}\eta\right)(\pi f_0)^{-2}.\label{def:tau2}
\end{eqnarray}
Note that $\tau_1$ and $\tau_2$ are simply the Newtonian and
post$^1$-Newtonian contributions to the time it takes for the carrier
gravitational wave frequency to evolve from $f_0$ to infinity.
The advantage of these variables is that the metric coefficients
in $(\tau_1,\tau_2)$ coordinates are
constant (in the limit $f_c \gg f_0$) for all templates.
This is because the phase of the waveform $\tilde u(f)$ is linear in the
Sathyaprakash variables, and so the integral in the definition of the
match (\ref{match definition}) depends only on the displacement
$(\Delta\tau_1,\Delta\tau_2)$ between the templates,
not on the location $(\tau_1,\tau_2)$ of the templates in the
dynamical parameter space.

The dynamical parameter-dependent
part of the templates' phase is given by
[Eq.~(\ref{2PN Psi}) truncated to first post-Newtonian order and
reexpressed in terms of $(\tau_1,\tau_2)$
using Eqs.~(\ref{def:tau1}) and (\ref{def:tau2})]
\begin{equation}
\label{1PN Psi}
\Psi (f;\tau_1,\tau_2) =
\frac{6}{5}\pi f_0 (f/f_0)^{-5/3}\tau_1
+ 2\pi f_0 (f/f_0)^{-1}\tau_2,
\end{equation}
and it is easy to read off $\psi_1$ and $\psi_2$ [Eq.~(\ref{psi-def})]
as the coefficients of $\tau_1$ and $\tau_2$ \cite{psi footnote}.
By inserting these $\psi_j$ into Eq.~(\ref{series functional}),
the relevant moment functionals can be expressed
in terms of the moments of the noise:
\begin{equation}
\label{1PN functionals}
\begin{array}{rcl}
{\cal J}[\psi_0] &=& 2\pi f_0~J(4), \\
{\cal J}[\psi_1] &=& 2\pi f_0~\frac{3}{5}J(12), \\
{\cal J}[\psi_2] &=& 2\pi f_0~J(10), \\
{\cal J}[\psi_0^2] &=& (2\pi f_0)^2~J(1), \\
{\cal J}[\psi_0\psi_1] &=& (2\pi f_0)^2~\frac{3}{5}J(9), \\
{\cal J}[\psi_0\psi_2] &=& (2\pi f_0)^2~J(7), \\
{\cal J}[\psi_1^2] &=& (2\pi f_0)^2~\frac{9}{25}J(17), \\
{\cal J}[\psi_1\psi_2] &=& (2\pi f_0)^2~\frac{3}{5}J(15), \\
{\cal J}[\psi_2^2] &=& (2\pi f_0)^2~J(13) .
\end{array}
\end{equation}

We can compute the needed moments of the noise by numerically evaluating the
integrals (\ref{moment definition}).
By setting the upper limit of integration to infinity,
i.e.\ by approximating $f_c/f_0$ as infinite
for all templates under consideration,
we find that the moments have the constant values given in Table~\ref{tab:J's};
and therefore the moment functionals (\ref{1PN functionals})
have the constant values given in Table~\ref{tab:cal J's}.
Inserting these values
into Eqs.~(\ref{premetric}) and (\ref{metric conversion}) yields,
for the coordinates
$(\lambda^0 = t_0,\;\lambda^1 = \tau_1,\;\lambda^2 = \tau_2)$,
the 3-metric and 2-metric
\begin{equation}
\gamma_{\alpha\beta} = (2\pi f_0)^2 \left(
\begin{array}{ccc}
+0.208 & -0.220 & -0.168 \\
 .     & +0.784 & +0.481 \\
 .     & .      & +0.309
\end{array}\right)\mbox{ and}
\end{equation}
\begin{equation}
\label{1st 1PN metric}
g_{ij} = (2\pi f_0)^2 \left(
\begin{array}{cc}
0.552 & 0.304 \\
 .    & 0.173
\end{array}\right)
\end{equation}
for the first LIGO noise curve, and
\begin{equation}
\gamma_{\alpha\beta} = (2\pi f_0)^2 \left(
\begin{array}{ccc}
+0.209 & -0.257 & -0.183 \\
 .     & +1.320 & +0.712 \\
 .     & .      & +0.407
\end{array}\right)\mbox{ and}
\end{equation}
\begin{equation}
\label{adv 1PN metric}
g_{ij} = (2\pi f_0)^2 \left(
\begin{array}{cc}
1.01 & 0.486 \\
 .   & 0.246
\end{array}\right)
\end{equation}
for the advanced LIGO noise curve
(where the dots denote terms obtained by symmetry).

We shall also estimate the errors in the metric coefficients
due to the approximation $f_c/f_0\to\infty$:
The moment integrals defined in Eq.~(\ref{moment definition}) can be
rewritten as
\[\int_{f_s/f_0}^\infty dx\frac{5x^{-q/3}}{x^{-4}+2(1+x^2)}
- \int_{f_c/f_0}^\infty dx\frac{5x^{-q/3}}{x^{-4}+2(1+x^2)},\]
where the first integral is the expression used in the above
metric coefficients and the second is the correction due to
finite $f_c/f_0$.
The second integral can be expanded to lowest order in $f_0/f_c$ as
\[\frac{5}{2(1+q/3)}(f_0/f_c)^{1+q/3},\]
and from this the errors in the moments
(and therefore in the metric coefficients)
due to approximating $f_c$ as infinite
are estimated to be
less than or of order ten percent for the first LIGO interferometers
and one percent for the advanced LIGO interferometers over most of
the relevant volume of parameter space.
Since the two-parameter, post$^1$-Newtonian continuous template family
is known to be inadequate for the task of searching for real binaries,
these errors are small enough to justify our use of the
$f_c\to\infty$ approximation in this exploratory analysis.

\subsection{Number of Search Templates Required}

Since the metric coefficients are constant in this analysis, the formula
for the required number of templates [Eq.~(\ref{cal N definition})] reduces to
\begin{equation}
\label{flat cal N}
{\cal N} = \frac{\sqrt{{\rm det}~\|g_{ij}\|}}{2(1-MM)} \int d\tau_1~d\tau_2 .
\end{equation}
The square root of the determinant of the metric is
given by $(2\pi f_0)^2~0.108$ for the advanced LIGO noise curve and
by $(2\pi f_0)^2~0.058$ for the initial LIGO noise curve,
so once we have decided on the range of parameters we deem astrophysically
reasonable we will have a formula for ${\cal N}$ as a function of $MM$.

The most straightforward belief to cherish about neutron stars is that they
all come with masses greater than a certain minimum $M_{\min}$, which
might be set to $0.2~M_\odot$
(based on the minimum mass
that any neutron star can have~\cite{Shapiro&Teukolsky})
or $1.0~M_\odot$
(based on the observed masses of neutron stars
in binary pulsar systems~\cite{Taylor}).
In terms of the variables $(M,\eta)$
the constraint $M_1 > M_{\min}$ and $M_2 > M_{\min}$ is easily expressed as
\[\frac{1}{2}M(1-\sqrt{1-4\eta}) > M_{\min},\]
but in terms of the Sathyaprakash variables [Eqs.~(\ref{def:tau1})
and (\ref{def:tau2})] the expression becomes rather
unwieldy to write down.
However, see Fig.~\ref{fig:volume} for a plot of
the allowed region in $(\tau_1,\tau_2)$ coordinates.

For this reason we have found it convenient to use a Monte Carlo integration
routine~\cite{numerical} to evaluate the coordinate volume integral
$\int d\tau_1~d\tau_2$.
The Monte Carlo approach becomes especially
attractive when evaluating the proper volume
integral $\int d\tau_1~d\tau_2~\sqrt{{\rm det}~\|g_{ij}\|}$ for cases where the
integrand is allowed to vary---and in fact may itself have to be evaluated
numerically, as will be the case for a post$^2$-Newtonian set of templates.
The integral has numerical values of $0.18$ and $24$ seconds$^2$
for initial and
advanced LIGO interferometer parameters, respectively, assuming a
$M_{\min}$ of $0.2~M_\odot$ and arbitrarily large $M_{\max}$.
The integral can be shown (numerically) to scale roughly as
$f_0^{-4.5}$ (independent of $f_0$)
and as $M_{\min}^{-2.7}$ for $M_{\min}$
ranging from 0.2 to 1.0 solar masses
(the dependence on $M_{\max}$ is negligible for any value greater than
a few solar masses).

Inserting the above numbers into Eq.~(\ref{flat cal N}), we find that
\begin{equation}
\label{1st cal N}
{\cal N} \simeq
2.7\times 10^5\left(\frac{MM}{0.03}\right)^{-1}
\left(\frac{M_{\min}}{0.2~M_\odot}\right)^{-2.7}
\end{equation}
for the first LIGO noise curve and
\begin{equation}
\label{adv cal N}
{\cal N} \simeq
8.4\times 10^6\left(\frac{MM}{0.03}\right)^{-1}
\left(\frac{M_{\min}}{0.2~M_\odot}\right)^{-2.7}
\end{equation}
for the advanced LIGO noise curve.
The fiducial value of $MM$ has been chosen as 0.97 to correspond to
an event rate of roughly 90 percent of the ideal event rate
[cf. Eq.~(\ref{ER vs MM})].
In terms of the template-spacing-induced fractional loss ${\cal L}$
of event rate, the number of templates required is
\begin{equation}
{\cal N}\simeq
2.4\times 10^5\left(\frac{{\cal L}}{0.1}\right)^{-1}
\left(\frac{M_{\min}}{0.2~M_\odot}\right)^{-2.7}
\end{equation}
for the first LIGO noise curve and
\begin{equation}
{\cal N}\simeq
7.6\times 10^6\left(\frac{{\cal L}}{0.1}\right)^{-1}
\left(\frac{M_{\min}}{0.2~M_\odot}\right)^{-2.7}
\end{equation}
for the advanced LIGO noise curve.

\subsection{Template Spacing}

With the aid of the metric coefficients given in
Eqs.~(\ref{1st 1PN metric}) and (\ref{adv 1PN metric}),
it is a simple task to select the locations of the templates
and the spacing between them.

Because the metric coefficients form a constant $2\times 2$ matrix,
we can easily find the eigenvectors $\bbox{e}_{x_1}$ and $\bbox{e}_{x_2}$
of $\| g_{ij} \|$
and use them as axes to lay out a grid of templates.
The numerical values are
\begin{equation}
\begin{array}{rcrcl}
\bbox{e}_{x_1} &=& 0.874\bbox{e}_{\tau_1}&+&0.485\bbox{e}_{\tau_2},\nonumber \\
\bbox{e}_{x_2} &=& -0.485\bbox{e}_{\tau_1}&+&0.874\bbox{e}_{\tau_2}
\end{array}
\end{equation}
for the first LIGO noise curve and
\begin{equation}
\begin{array}{rcrcl}
\bbox{e}_{x_1} &=& 0.899\bbox{e}_{x_1}&+&0.437\bbox{e}_{\tau_2},\nonumber \\
\bbox{e}_{x_2} &=& -0.437\bbox{e}_{x_1}&+&0.899\bbox{e}_{\tau_2}
\end{array}
\end{equation}
for the advanced LIGO noise curve.
The infinitesimal proper distance is given in terms of the
eigen-coordinates as $E_1(dx_1)^2+E_2(dx_2)^2$,
where $E_1$ and $E_2$ are the eigenvalues of the metric.

Therefore we simply use Eq.~(\ref{MM vs dl})
to obtain the template spacings
\begin{equation}
dx_j = \sqrt{\frac{2(1-MM)}{E_j}},\quad j=1,2 .
\end{equation}
We find that the eigenvalues of the metrics
(\ref{1st 1PN metric}) and (\ref{adv 1PN metric}) are
$(2\pi f_0)^2$ times 0.721 and 0.00427 (first LIGO),
and $(2\pi f_0)^2$ times 1.25 and 0.00984 (advanced LIGO).
Therefore the template spacings are given by
\begin{eqnarray}
dx_1 &=& 0.22\mbox{ ms}
\left(\frac{1-MM}{0.03}\right)^{1/2}
\left(\frac{f_0}{200~{\rm Hz}}\right)^{-1},\nonumber \\
dx_2 &=& 2.9\mbox{ ms}
\left(\frac{1-MM}{0.03}\right)^{1/2}
\left(\frac{f_0}{200~{\rm Hz}}\right)^{-1}
\end{eqnarray}
for the first LIGO noise curve and by
\begin{eqnarray}
dx_1 &=& 0.49\mbox{ ms}
\left(\frac{1-MM}{0.03}\right)^{1/2}
\left(\frac{f_0}{70~{\rm Hz}}\right)^{-1},\nonumber \\
dx_2 &=& 5.6\mbox{ ms}
\left(\frac{1-MM}{0.03}\right)^{1/2}
\left(\frac{f_0}{70~{\rm Hz}}\right)^{-1}
\end{eqnarray}
for the advanced LIGO noise curve.
Figure~\ref{fig:contours} shows the locations of some possible templates
superposed on a contour plot of the match with the template in the center
of the graph.

\subsection{Computing Power Requirements}

Drawing on the previous work of Schutz~\cite{Schutz}
concerning the mechanics of fast-Fourier-transforming
the data, we can estimate the CPU power required
to process the interferometer output on-line through a single-pass
(non-hierarchical) search
involving ${\cal N}$ templates.

Although the data will be sampled at a rather high rate (tens of kHz),
frequencies above some upper limit
$f_u\simeq 4f_0$ can be thrown away
(in Fourier transforming the data) with only negligible
effects on the signal-to-noise ratio.
This lowers the effective frequency of sampling to $2f_u$
(the factor of two is needed so that the Nyquist frequency is $f_u$),
and thereby
considerably reduces the task of performing the inner-product integrals.
If the length of the array of numbers required to store a template is $F$
and that required to store a given stretch of data is $D$,
the number of floating point operations required to process that data stretch
through ${\cal N}$ filters is
\begin{equation}
\label{N operations}
N_{\rm f.o.} \simeq D{\cal N}(16+3{\rm log_2}F)
\end{equation}
[cf. Eq.~(16.37) of Schutz~\cite{Schutz}, with
the fractional overlap between data segments $x$ chosen as roughly $1/15$].

Actually, $F$ varies from filter to filter, but most of the
search templates occupy regions of parameter space where the mass is
very low---and thus the storage size of the template,
\begin{equation}
\label{storage size}
F \simeq 2f_u \tau_1(f_0/f_s)^{8/3},
\end{equation}
is very large \cite{filter length}.
The longest filter is the one computed for two stars of mass $M_{\min}$,
so by inserting $\eta=1/4$ and $M=2M_{\min}$
into Eq.~(\ref{def:tau1}) and combining with Eq.~(\ref{storage size}),
we find that we can make a somewhat pessimistic estimate
of the computational cost by using
\begin{equation}
F\simeq 2^{14}\left(\frac{M_{\min}}{0.2~M_\odot}\right)^{-5/3}
\left(\frac{f_s}{10~{\rm Hz}}\right)^{-8/3}
\left(\frac{f_0}{70~{\rm Hz}}\right).
\end{equation}

The required CPU power ${\cal P}$ for an on-line search
is obtained by dividing Eq.~(\ref{N operations}) by
the total duration of the data set,
\[T_{\rm tot}\simeq D/(2f_u),\]
to find that
\begin{equation}
\label{CPU power}
{\cal P}\simeq 2{\cal N}f_u(16+3{\rm log_2}F).
\end{equation}
Combining Eq.~(\ref{CPU power})
with Eqs.~(\ref{1st cal N}) and (\ref{adv cal N}) gives us
\begin{equation}
{\cal P}\simeq
20~{\rm Gflops}~\left(\frac{{\cal L}}{0.1}\right)^{-1}
\left(\frac{M_{\min}}{0.2~M_\odot}\right)^{-2.7}
\left(\frac{f_0}{200~{\rm Hz}}\right)^{-1.5}
\end{equation}
for an on-line search by the first LIGO interferometers and
\begin{equation}
{\cal P}\simeq
270~{\rm Gflops}~\left(\frac{{\cal L}}{0.1}\right)^{-1}
\left(\frac{M_{\min}}{0.2~M_\odot}\right)^{-2.7}
\left(\frac{f_0}{70~{\rm Hz}}\right)^{-1.5}
\end{equation}
for the advanced LIGO interferometers.

Although the estimates in the paragraph above are not to be believed beyond
a factor of order unity, the magnitude of the numbers shows that a
hierarchical search strategy may be desirable to keep the computing power
requirements at a reasonable level for non-supercomputing facilities.
That is, the data would first be filtered through a more widely spaced
(low minimal match) set of
templates with a relatively low signal-to-noise threshold, and only the
segments which exceed this preliminary threshold would be analyzed with the
finely spaced (high minimal match) templates.

The metric-based formalism of this paper only holds for the finely spaced set
of templates used in the final stage of the hierarchical search; the template
spacing used in the earlier stages of the search will have to be chosen using
more complex methods
such as those of Sathyaprakash and Dhurandhar~\cite{S&D 1,S&D 2}.

\subsection{Comparison With Previous Results}

The only previous analysis of the problem of choosing the discrete
search templates from the two-parameter, restricted
post$^1$-Newtonian continuous template family
is that of Sathyaprakash~\cite{Sathya note}, in which he found that
the entire volume of parameter space corresponding to
$M_{\min} = 1~M_\odot$ could be covered by a set of templates which vary
only in $\tau_1 + \tau_2$---thereby reducing the effective dimensionality
of the mass parameter space to one.
This implied a value of ${\cal N}$
similar to that obtained in the one-parameter (Newtonian template) analysis
of Dhurandhar and Sathyaprakash in Ref.~\cite{S&D 2}.

It is not possible to fairly compare my value for ${\cal N}$
to the values given by Dhurandhar and Sathyaprakash
in Table II of Ref.~\cite{S&D 2}
due to our differing assumptions concerning the sources
and the desirable level of the minimal match.
Therefore I will compare the assumptions.

Dhurandhar and Sathyaprakash typically consider a minimal match of 0.8 or 0.9
rather than 0.97,
This would lead to a loss of thirty to fifty percent of the ideal event rate
[cf.\ Eq.~(\ref{ER vs MM})].
If the current ``best estimates'' of inspiraling binary event rates
\cite{Phinney,Narayan et al} are correct,
the ideal event rate for LIGO and VIRGO
will not be more than about one hundred per year
even when operating at the ``advanced interferometer'' noise levels,
and the loss of up to half of these events would be unacceptable.

{}From Eqs.~(\ref{1st cal N}) and (\ref{adv cal N}) it can be seen that
the dependence of ${\cal N}$ on $M_{\min}$ is the most important factor
influencing the number of templates.
The two-parameter analysis of Sathyaprakash \cite{Sathya note}
uses a value for $M_{\min}$ of $1~M_\odot$, which is based on
the statistics of (electromagnetically-) known binary pulsars.
However, because there is no known, firm physical mechanism
that prevents neutron stars from forming with masses between 0.2 and 1
$M_\odot$, LIGO and VIRGO should use a discrete template family with
$M_{\min}=0.2~M_\odot$.
After all, laser interferometer gravitational wave detectors are expected
to bring us information about astronomical objects as yet unknown.

During the final stages of completion of this manuscript, a new preprint by
Balasubramanian, Sathyaprakash, and Dhurandhar appeared in the xxx.lanl.gov
archive~\cite{BSD}, applying differential geometry to the problem of
detecting compact binary inspiral events and extracting source parameters
from them.
The preprint applies the tools of differential geometry primarily to the
problem of parameter measurement rather than that of signal detection,
and so does not develop the geometrical formalism as far
as is done in Sec.\ \ref{sec:formalism} of this paper.
The metric constructed in Ref.~\cite{BSD}
is identical to the information matrix
which was suggested for use in the construction of a closely-spaced
discrete template family in the authors' previous work
\cite{S&D 2}.
While this is quite useful for parameter measurement,
it neglects maximization over kinematical parameters
and thus is not very useful for the construction of search templates.
Also, the assumptions about $MM$ and $M_{\min}$ are no different from
those made in previous work up to and including Ref.~\cite{Sathya note},
and so the result for ${\cal N}$ is no different.

The main difference between the results of Ref.~\cite{BSD}
and previous analyses by the same authors---and therefore
the most important part of the preprint as far as the detection
problem is concerned---is the introduction of the
possibility of choosing search templates to lie outside the manifold
of the continuous template family.
Using an {\it ad hoc} example,
the authors show that such a placement of templates can result in
a spacing roughly double that between discrete templates chosen from the
manifold formed by the continuous template family.
My analysis in this paper does not consider this possibility,
but the formalism of Sec.\ \ref{sec:formalism} can easily be extended
to investigate this problem in the future.

\section{Conclusions}
\label{sec:conclusion}

\subsection{Summary of Results}

This paper has presented
a method for semi-analytically calculating the number of
templates required to detect gravity waves from inspiraling binaries
with LIGO as a function of the fraction of event rate lost
due to the discrete spacing of the templates in the binary parameter space.
This method, based on differential geometry,
emphasizes that ultimately a finer template spacing is required than
has previously been taken as typical in the literature, in order to retain
a reasonable fraction of the event rate.
This paper details the first calculation of this kind
that uses post-Newtonian templates and a noise curve which takes into
account the coloration of noise in the detector due to both standard
recycling photon shot noise and thermal noise in the suspension of the
test masses.

The result is that it is possible to search the data
for binaries containing objects more massive than $0.2~M_\odot$
thoroughly enough to lose only $\sim$ 10 percent of the ideal event rate
without requiring a quantum leap in computing technology.
The computational cost of such a search,
conducted on-line using a single pass through the data,
is roughly 20 Gflops for the first LIGO interferometers (ca.\ 2000)
and 270 Gflops for the advanced LIGO interferometers (some years later).
This is feasible (or very nearly feasible)
even for a present-day supercomputing facility,
but a hierarchical search strategy
(using as its first stage a widely-spaced
set of templates similar to that analyzed by Sathyaprakash~\cite{Sathya note})
may be desirable to reduce the cost.

\subsection{Future Directions}

A thorough investigation of hierarchical search strategies is in order:
How should the threshold and the minimal match of the first stage be set in
order to minimize the CPU power required while keeping the false alarm and
false dismissal rates at acceptable levels?
How would non-Gaussian noise statistics affect the first stage threshold and
minimal match?
Would a hierarchical search benefit by using more than two stages?
How is the threshold affected by the minimal match when the approximation
of high signal-to-noise ratio can no longer be made?

The formalism of this paper should be applied to choose discrete templates
from a better continuous template family than the one considered here.
The best two-parameter templates will be based on the highest
post-Newtonian order computations that have been performed for
circularized, spinless binaries,
augmented perhaps by terms of still higher order from
the theory of perturbations of Schwarzschild or Kerr spacetime.
I plan to soon apply my geometric formalism to the post$^{2.5}$-Newtonian
templates which are currently the best available.
The areas of parameter space where spins cannot be neglected
(noted by Apostolatos in Ref.~\cite{Apostolatos}) must also be investigated,
and the inclusion of an orbital eccentricity parameter should be considered.

There are several more issues which I plan to address using my formalism
or some extension of it.
An analysis needs to be made for the case when the signal is not
identical to some member of the continuous template family
(i.e.\ the fitting factor is not equal to one);
and the result of such an analysis should be used
to set definite goals for both the fitting factor and the minimal match
in terms of event rate.
The effect of non-quadrupolar harmonics of the gravitational wave on the
construction of search templates should be considered.
These harmonics have been ignored
in all previous analyses of detection and even of parameter measurement,
but they may have a noticeable effect
when a very high minimal match is desired.
Finally, a systematic investigation of the optimal choice of search templates
outside the continuous template family is in order.
This problem has been addressed in a preliminary way in Ref.~\cite{BSD},
but is deserving of further scrutiny.

\acknowledgments

My thanks to Theocharis Apostolatos and Eanna Flanagan
for helping me get started.
I am most indebted to Kip Thorne
for his guidance and his patience in reviewing the manuscript.
This work was supported in part by my NSF graduate fellowship
and in part by NSF grant PHY-9424337.

\begin{figure}
\caption{A schematic depiction of the manifold formed by the continuous
template family, here represented as a two-dimensional surface lying within
a three-dimensional space.
The discrete template family, shown by the dots, resides
within this manifold.
The $\times$ indicates the location of an actual signal, which because it is an
exact solution to the Einstein equations does not lie within the manifold.
The + marks the spot in the manifold which is closest to
(has the highest inner product with) the signal.
In general, this location falls in between the actual discrete templates.
}
\label{fig:manifold}
\end{figure}

\begin{figure}
\caption{The two-dimensional region of parameter space inhabited by binaries
composed of objects more massive than 1 $M_\odot$.
$\tau_1$ and $\tau_2$ are expressed (in milliseconds) for $f_0 = 200$ Hz,
but the shape of the region does not change with $f_0$.
The upper boundary of the wedge is set by $M_{\min}$.
The left-hand boundary is set by $M_{\max}$, but is essentially identical
to the $\tau_2$ axis for $M_{\max}$ greater than a few solar masses.
The region below the wedge corresponds to $\eta > 1/4$,
which is {\it a priori} impossible.
}
\label{fig:volume}
\end{figure}

\begin{figure}
\caption{Locations of various discrete templates for the first LIGO
interferometers are shown by dots.
The contours indicate the value of the match between a template located
at $(\tau_1,\tau_2)$ and the template located in the center of the figure.
The contours are drawn at match values of 0.97, 0.98, and 0.99.
The semimajor and semiminor axes of the contour ellipse (along which
the templates are placed) do not appear to be perpendicular
because of the aspect ratio used to make the graph readable.
}
\label{fig:contours}
\end{figure}

\begin{table}
\caption{Numerical values of the moments of the noise in the limit
$f_c/f_0\to\infty$, for the noise curves of the first LIGO interferometers
($f_s/f_0 = 1/5$) and advanced LIGO interferometers ($f_s/f_0 = 1/7$).
}
\begin{tabular}{ccc}
Noise moment & First LIGO value & Advanced LIGO value \\
\tableline
J(1) & 1.27 & 1.26 \\
J(4) & 0.927 & 0.919 \\
J(7) & 1 (exact) & 1 (exact) \\
J(9) & 1.24 & 1.26 \\
J(10) & 1.44 & 1.49 \\
J(12) & 2.13 & 2.31 \\
J(13) & 2.69 & 3.03 \\
J(15) & 4.67 & 5.80 \\
J(17) & 8.88 & 12.7 \\
\end{tabular}
\label{tab:J's}
\end{table}

\begin{table}
\caption{Numerical values of the moment functionals, under the same
assumptions as in Table I.}
\begin{tabular}{ccc}
Moment functional & First LIGO value & Advanced LIGO value \\
\tableline
${\cal J}[\psi_0]$ & $(2\pi f_0)$0.927 & $(2\pi f_0)$0.919 \\
${\cal J}[\psi_1]$ & $(2\pi f_0)$1.28 & $(2\pi f_0)$1.38 \\
${\cal J}[\psi_2]$ & $(2\pi f_0)$1.44 & $(2\pi f_0)$1.49 \\
${\cal J}[\psi_0^2]$ & $(2\pi f_0)^2$1.27 & $(2\pi f_0)^2$1.26 \\
${\cal J}[\psi_0\psi_1]$ & $(2\pi f_0)^2$0.743 & $(2\pi f_0)^2$0.756 \\
${\cal J}[\psi_0\psi_2]$ & $(2\pi f_0)^2$ (exact) & $(2\pi f_0)^2$ (exact)\\
${\cal J}[\psi_1^2]$ & $(2\pi f_0)^2$3.20 & $(2\pi f_0)^2$4.56 \\
${\cal J}[\psi_1\psi_2]$ & $(2\pi f_0)^2$2.80 & $(2\pi f_0)^2$3.48 \\
${\cal J}[\psi_2^2]$ & $(2\pi f_0)^2$2.69 & $(2\pi f_0)^2$3.03 \\
\end{tabular}
\label{tab:cal J's}
\end{table}


\begin{references}
\bibitem{phase footnote} Note that the $\Phi_0$ used here is the
coalescence phase of the {\it waveform}.
In the restricted post-Newtonian approximation, this is twice the
coalescence phase of the {\it orbit} $\phi_0$, which is used
by Cutler and Flanagan in Ref.~\cite{Cutler&Flanagan}.
\bibitem{gamma footnote} This $(N+1)$-dimensional metric
$\gamma_{\alpha\beta}$ is related to the Fisher information matrix, which
was used by Dhurandhar and Sathyaprakash
in the appendix of Ref.~\cite{S&D 2} to compute the template
spacing analytically in the small-spacing limit.
However, in taking the absolute value to maximize over $\Delta\Phi_0$, we
have lost some dependence
on the dynamical (intrinsic) template parameters,
so our $\gamma$-metric is not even
the projection of the information matrix onto
the subspace perpendicular to $\Phi_0$.
This lets us tolerate
a greater mismatch between the templates' dynamical parameters
than was possible before maximization,
thereby reducing the total number of templates required.
\bibitem{psi footnote} The $\psi_j$ given here
are different from those given in Eq.~(13a-3) of Ref.~\cite{Sathya note}
because Sathyaprakash does not use $t_0$ and $\Phi_0$ as kinematical
parameters,
but rather the time and phase at which the quadrupole part of the waveform
reaches a frequency of $f_0$.
\bibitem{filter length} The storage size of the template
is equal to the effective sampling frequency $2f_u$ times the duration
of the template waveform.
In the restricted post-Newtonian approximation,
the duration of a binary chirp waveform template is equal to
the time it takes for the gravitational wave frequency
to sweep from the seismic frequency $f_s$ up to infinity.
Since $\tau_1\propto f_0^{-8/3}$
is the dominant (Newtonian) contribution to the time it takes the
gravitational wave frequency to sweep from $f_0$ up to infinity,
the duration of a waveform template is roughly
$\tau_1 (f_0/f_s)^{8/3}$.
Thus we obtain Eq.~(\ref{storage size}).
\end{references}
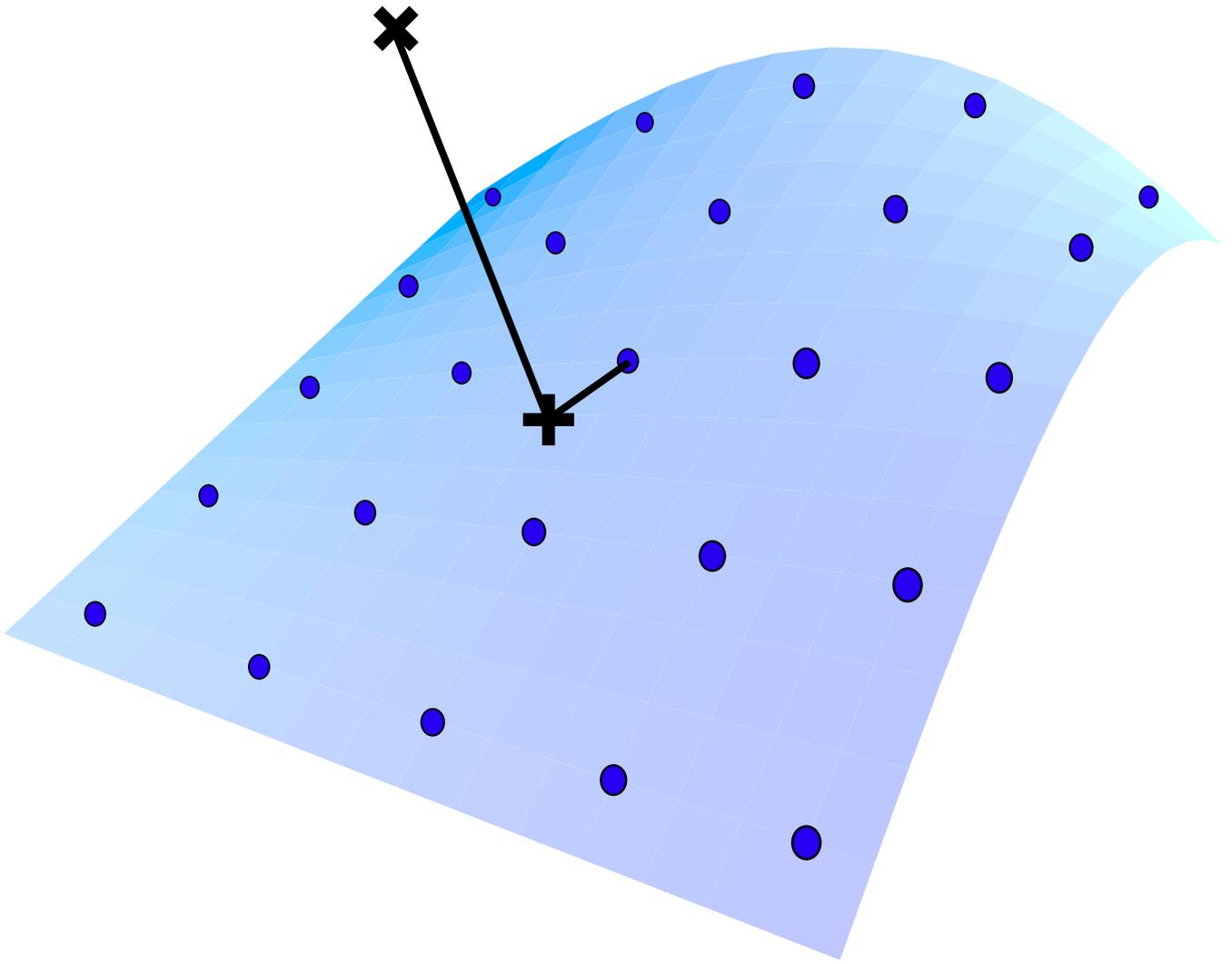
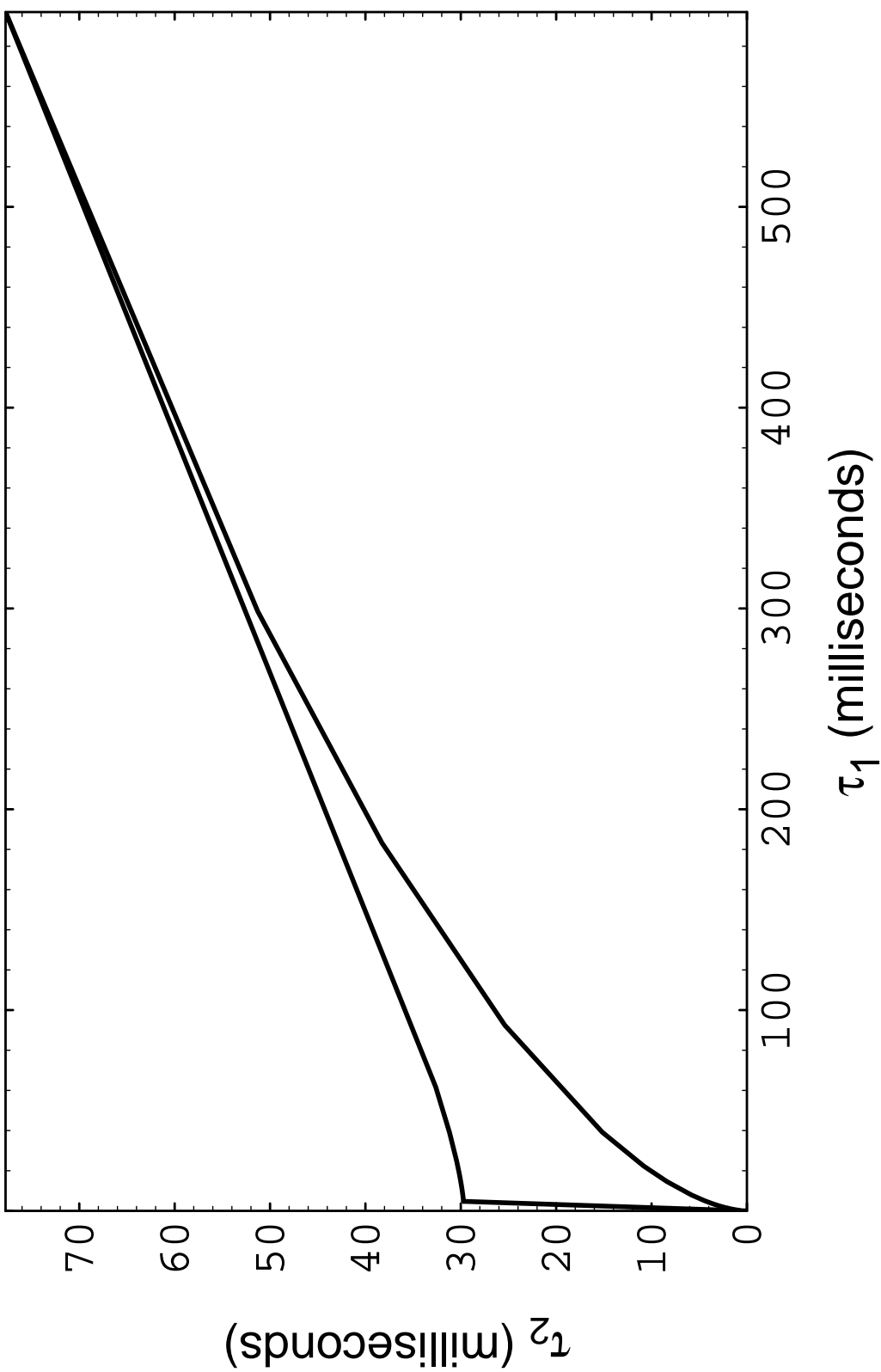
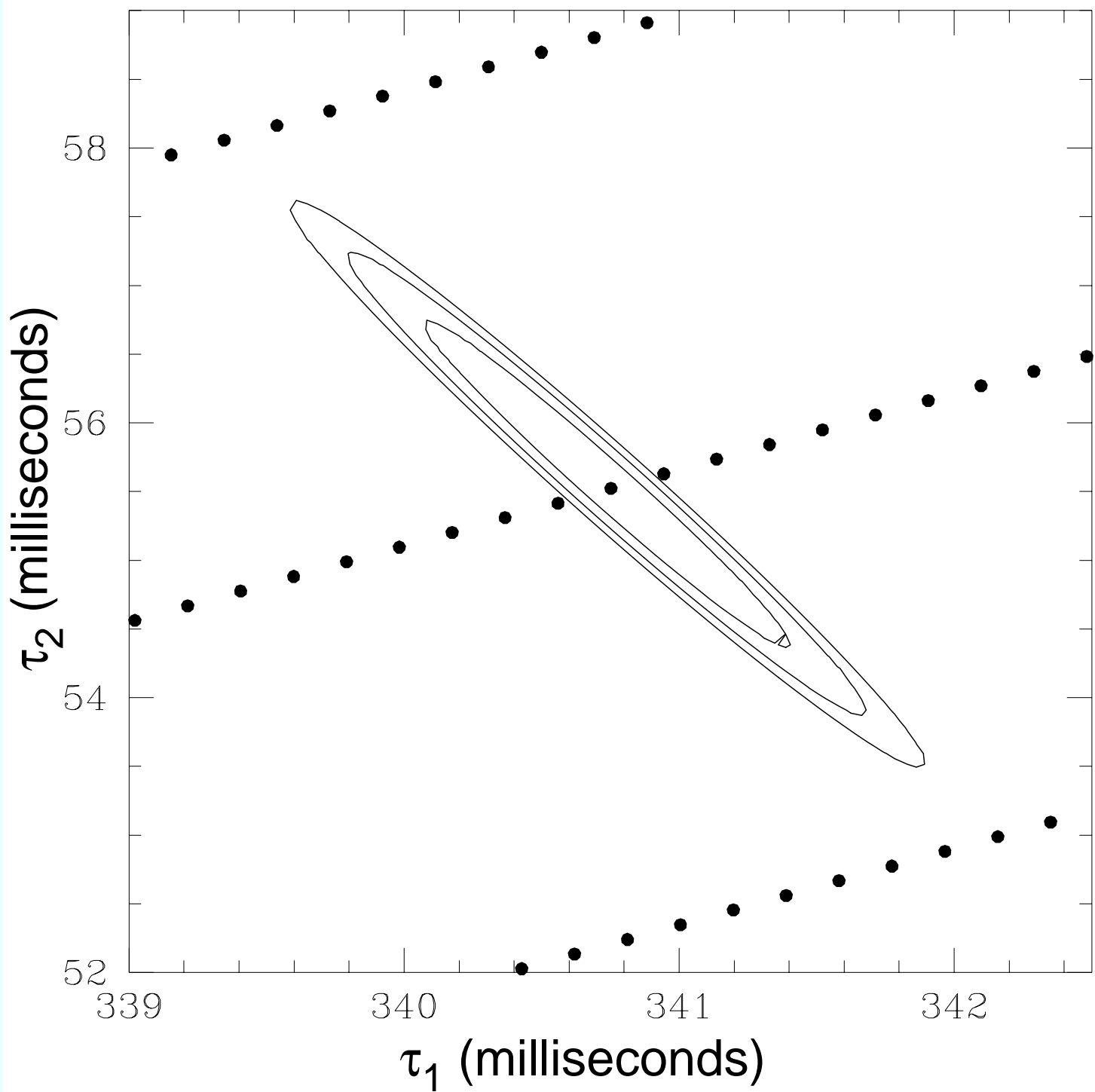
\end{document}